# Gap Plasmon Polariton Structure for Very Efficient Micro to Nano Scale Interfacing

Pavel Ginzburg, David Arbel [*], and Meir Orenstein

*Department of Electrical Engineering, Microphotonics lab., Technion – Israel Institute of*

*Technology,*

*Haifa 3200, Israel*

The seamless transition between micro-scale photonics and nano-scale plasmonics requires the mitigation between different waveguiding mechanisms as well as between few orders of magnitude in the field lateral size, down to a small fraction of a wavelength. By exploiting gap plasmon polariton waves both at the micro and nano scale, very high power transfer efficiency (>60%) can be achieved using an ultrashort (few microns) non adiabatic tapered gap plasmon waveguide. Same mechanism may be used to harvest impinging light waves and direct them into a nano hole or slit, to exhibit an anomalous transmission - without the conventional periodic structures. The special interplay of plasmonic and oscillating modes is analyzed.

PACS numbers: 73.20.Mf, 42.82.m, 81.07.b

Nano-Photonics has gained much interest in recent years, with newly developed nano-fabrication techniques allowing the realization of sub-micron optical devices, such as high-index-contrast channel waveguides, and Photonic-Crystal waveguides [1],[2]. Yet, the micron-scale wavelength of light limits the miniaturization of optical devices. Surface Plasmon Polariton (SPP) optical waves on a metal-dielectric interface can be confined much below the optical wavelength [3], and offer a route to overcome the diffraction-limit. Research of sub-wavelength SPP optics shows promise for the realization of nanometer-size Photonic Integrated Circuits, for applications such as optical interconnects, signal processing and nano-sensing [4]. Metal stripes tens of nanometer thick, surrounded by a dielectric cladding, can potentially serve as low loss SPP waveguides, as was shown theoretically [5],[6] and experimentally [7],[8]

A major concern of basic interest and for practical implementation of plasmonic circuits, is achieving efficient interfacing between conventional micron-size photonic waveguides to tens-of-nm-size plasmon carrying waveguides. A closely related field of interest is the efficient transmission of light through nano-holes and slits in metal layers [9], where the micron-scale wavelength impinging field is harvested employing a periodic structure, and then transmitted through the nano hole [10],[11]. Current schemes for coupling of light to a gold stripe plasmon waveguide encompasses end-fire excitation from a single-mode optical fiber [7],[8], and side coupling between SOI waveguides and silver plasmon waveguides [12]. While a significant portion of previous research was focused on slab and stripe plasmon waveguides consisting of a metal core with a dielectric cladding ([4]-[8] and ref. therein), we studied a plasmon waveguide with the inverted structure of a dielectric core with a metal cladding [13], coined Plasmon Gap Waveguide (PGW) [14]. The PGW provides tight optical confinement by the metallic cladding, and should be better coupled to a dielectric waveguide due to the continuity of the waveguide core, as well as to regular plasmon waveguides by proper plasmonic coupling schemes.

In this letter we investigate and analyze the interplay and coupling of modes in a tapered PGW structure. A very short taper exhibits surprisingly high efficiency plasmon assisted coupling from a micron-size Silica/Air-clad dielectric waveguide to a 50nm Silica/Gold-clad Plasmon Gap Waveguide. Power transfer efficiency of more than 70% is exhibited. Some of the fundamental findings described here are: the cut-off of a dielectric waveguide mode after which it changes into a plasmon surface wave in PGW; an interpretation for the lack of adiabaticity constraints for efficient PGW tapers; and a novel highly efficient method for harvesting light into nano slits – not using a periodic structure – namely funnel shape metal interfaces, that can be used to transmit "more than 100%" via nano holes.

The dispersion relations of Plasmon Gap Waveguides, consisting of a dielectric core with a metal cladding were discussed in different contexts by Economou [15], and Kaminow et. al. [16]. Prade et. al. [17] analyzed theoretically a loss-less waveguide with a dielectric core and a cladding with a negative real dielectric constant. The guided modes can be either Plasmon modes, of exponential nature, or Oscillating modes with a trigonometric behavior similar to modes of a dielectric waveguide. Recent works have analyzed the trade-off between confinement and propagation losses in a PGW [18], including the effects of empirical metal data [19]. We would like to emphasize an important difference between the plasmonic modes of slab and gap structures. While for a gap structure plasmonic modes exist both for nanoscale and microscale waveguide dimension, an essential characteristic for the desired seamless transition, this has no counterpart in a metal slab plasmon waveguide. As the fields quickly decay in the metal core for slab thickness larger than the skin depth (~100nm for the materials and wavelength considered here), slab plasmon modes split into single interface SPP modes.

The PGW we have analyzed consists of a planar Silica slab of width $2 \cdot d$ along the X-axis surrounded on both sides by a semi-infinite Gold cladding, with the waves propagating along the Z axis. We chose the telecom wavelength of 1.55μm, which has the advantage of exhibiting lower metal losses than in the visible. The values used for the dielectric constants were: $\varepsilon_m(Gold) = -132 - i \cdot 12.6$, and $\varepsilon_d(SiO_2) = 2.085$ [20], with the complex dielectric constant of Gold taking into account metal loss. The PGW can have oscillating TE-polarized modes, but we will focus on TM-polarized modes, for which plasmon modes exist as well, with $\vec{E} = \hat{x} \cdot f(x) \cdot e^{-i \cdot \beta \cdot z}$ and $\vec{H} = \hat{y} \cdot g(x) \cdot e^{-i \cdot \beta \cdot z}$, where $\beta = \beta_R - i \cdot \beta_I$ is the complex propagation constant.

The dispersion relations for the two families of symmetric guided modes are [17]:

$$\text{Plasmon}: \quad tgh(k_p d) = -(k_m/\varepsilon_m)/(k_p/\varepsilon_d) \quad (1a)$$

$$\text{Oscillating}: \quad tg(k_O d) = (k_m/\varepsilon_m)/(k_O/\varepsilon_d) \quad (1b)$$

where $k_p^2 = \beta^2 - k_0^2 \varepsilon_d$ ; $k_O^2 = k_0^2 \varepsilon_d - \beta^2$ ; $k_m^2 = \beta^2 - k_0^2 \varepsilon_m$.

(For the asymmetric modes dispersion relations, the (tan) should be replaced by (-cot)).

The solution of the complex dispersion relations yields $\beta_R$ and $\beta_R$ vs. the gap width (which we intend to linearly vary), plotted in Figs. 1a and 1b respectively. In Fig. 1a, the Plasmon and Oscillating modes are separated by the core light-line, with $\beta_{R,plasmon} > k_0 \cdot n_d$, and $0 < \beta_{R,oscillating} < k_0 \cdot n_d$.

Two plasmon mode solutions exist: a symmetric mode for any gap width, and an anti-symmetric mode that has a cut-off for gap widths below $\sim \lambda_0/2n_d$, making the symmetric plasmon mode a single-mode of the structure in the nano regime. Odd and Even oscillating modes exist as well in the gap plasmon waveguide, each has a cut-off width, below which it becomes evanescent with very high losses, as seen in Fig. 1b. An interesting feature found here is that the first anti-symmetric plasmon mode, $TM_1$, for a gap width below its cut-off value, transforms into the first Odd Oscillating mode. Thus this oscillating mode exhibits two cutoff values, from below – into an evanescent mode and from above – into plasmon mode. As no additional plasmon modes exist below $\beta_{R,Plas}(d=\infty)$, the single-surface plasmon-polariton propagation constant, all the other Oscillating modes approach the light-line asymptotically, but do not cross it.

The first stage of harvesting the light from the micro-scale Silica/Air dielectric waveguide calls for an efficient conversion of the optical mode to a plasmonic mode of interest, supported by a PGW (Fig. 2a). Due to the symmetry of the structure and excitation, the modes excited at the PGW will be symmetrical as well, i.e. the relevant modes for our analysis are the symmetrical Plasmon mode $TM_0$, and the lowest order even Oscillating mode $TM_2$. The modes at the input plane of the PGW are plotted in Figs. 2b-2d - Plasmon, Oscillating, and dielectric input waveguide modes respectively.

We analyzed in detail and closed form the transmission and reflection at the interface between the dielectric and plasmon waveguides, where the incoming mode of the dielectric waveguide couples into both plasmon and oscillating modes of the PGW, as well as back reflected and couples to guided and radiation modes of the dielectric slab. The width of the dielectric waveguide was scanned in the range of 1.05 to 1.5 μm, limited from below by the cut-off of the PGW $TM_2$ mode (we intentionally analyzed the non ideal case where light can partially couple also to unwanted modes). It is evident from Fig. 3a that at a waveguide width of 1.25μm almost no energy is back reflected in the photonic-plasmonic conversion process, and as much as 85% of the incident photonic power is converted to the plasmon mode of the PGW, while 15% is coupled to the Oscillating mode.

The second stage of harvesting is the collection of the power converted to the plasmonic mode, and guiding it with minimal losses into the nano waveguide, 50nm wide in our example. A linearly tapered structure is employed to perform this task. The width change along the Z-axis facilitates also some coupling between the plasmon and oscillating mode in the PWG. In our analysis we dwell with this inter-mode coupling, and optimize the taper angle as to maximize plasmon mode transmission to the 50nm narrow end.

The linearly tapered PGW was discretized to a sequence of short straight segments with varying width, and we employed a Transfer Matrix formalism to calculate the amplitudes of the forward and backward modes at each segment. Modes coupling coefficients were calculated by overlap integrals of the form:

$$C_{i,j}^{L,R} = \langle \varepsilon_i^L | h_j^R \rangle = \iint_{plane\ z=0} \varepsilon_i^L \times h_j^R \cdot dx \quad (2)$$

where $\varepsilon_i^L$ is the electric field of mode i in the Left segment, and $h_j^R$ is the magnetic field of mode j in the right segment, and i, j are plasmon / oscillating forward / backward modes. We did not assume orthogonality of the modes at each segment, and coupling coefficients within the segment $C_{i,j}^{L,L}$, $C_{i,j}^{R,R}$ were calculated and not assumed to be zero. This scheme fits well lossy geometries and also stabilizes the numerical scheme. Assuming mode orthogonally, the zero elements of the matrices accumulate numerical noise and grow over long propagation distances while for nonorthogonity these close-to-zero orthogonal matrix elements stay ~12 orders of magnitude below the significant amplitudes.

The coupling of plasmon and Oscillating modes along the tapered waveguide was calculated and is exhibited by the power oscillation of the modes (Fig. 3b). The optimal design, obtained for a taper angle of 5.5 degrees (Fig 4), is a result of a tradeoff between the point of maximal coupling and higher propagation losses originating from longer taper lengths. As shown in Fig. 3b, the cutoff point of the oscillating mode is achieved approximately where the coupled power is maximal at the plasmonic mode. This taper angle results in a very short 6μm and very nonadiabatic funnel, an order of magnitude shorter than the common wisdom in regular photonics circuits. The two fundamental reasons for not requiring adiabaticity from plasmonic gap tapers are the dominance of propagation loss and the fact that there are no other forward propagation modes to couple to, neither guided nor radiation. This is in contrast to a dielectric taper, where the primary mechanism of non adiabatic taper loss is the coupling to forward propagation radiation modes. The evolution of the single-mode plasmon mode beyond the cut-off of the Oscillating mode exhibits very small coupling to the backward plasmon mode, as can be seen in Fig. 3b. More than 70% of the power incident from the dielectric waveguide is coupled eventually into the plasmon mode of the 50nm wide PGW. To verify this result, finite difference time domain (FDTD) simulations of the tapered PGW coupler were performed and confirmed the results of the transfer matrices analysis, exhibiting a high coupling efficiency of ~60%. Switching the input field from TM to TE results in an almost complete shut-down of the transmission with all energy reflected back, as seen in Fig. 5. This validates that the high coupling efficiency is achieved through coupling to Plasmon surface waves.

This high harvesting efficiency from a micron-scale input field into a 50nm plasmon gap can also be used for anomalous transmission experiments through nano-sized holes or slits in thin metal films, similar to those describes in [9], [10]. In the latter, harvesting enhancement is achieved by a periodic array of grooves surrounding the slit, used to couple the impinging field into surface plasmon polariton. The method described in our letter of using a funnel structure offers roughly an order-of-magnitude improvement over the theoretical calculations of periodic arrays, while not requiring operating at a resonance wavelength of the structure. In addition, for input fields with lateral beam dimensions of about micron, the periodic structure is almost inapplicable as the number of periods in the beam cross section is vanishingly small.

In Summary, the analysis predicts a coupling loss of less than 2dB from a micron-size dielectric waveguide, to a 50nm wide plasmon gap waveguide, by using an ultra short (6 μm) optimized tapered plasmon waveguide coupler. Thus it offers a promising method to couple micro-optics circuits to nano plasmon-optics devices, and a step towards large scale integrated photonic circuits.

[*]Corresponding author.

Email address: arbeld@tx.technion.ac.il

[1]  R. Nagarajan et. al., IEEE J. Sel. Top. In Quantum Elec. 11, 50-65 (2005).

[2]  R. Baets, Tutorial in ECOC 2003 conference, Rimini, Italy (2003).

[3]  H. Raether, "Surface Plasmons", Springer Tracts in Modern Physics, Vol. 111, Springer-Verlag, Berlin (1988).

[4]  W. L. Barnes et. al, Nature 424, 824-830 (2003).

[5]  J. J. Burke et. al., Phys. Rev. B 33, 5186–5201 (1986).

[6]  P. Berini, Phys. Rev. B 61, 10484–10503 (2000).

[7]  R. Charbonneau et al, Opt. Lett. 25, 844 (2000).

[8]  T. Nikolajsen et. al., Appl. Phy. Lett. 82, 668-670 (2003).

[9]  T. W. Ebbesen et al., Nature (London) **39**1, 667 (1998); H. F. Ghaemi et al., Phys. Rev. B **5**8, 6779 (1998).

[10] F. J. Garcia-Vidal et. al., Phys. Rev. Let. 90, 213901-1 (2003).

[11] H. J. Lezec, and T. Thio, Optics Exp. 12, 3629-3651 (2004).

[12] M. Hochberg et. al., Opt. Exp. **12**, 5481-5486 (2004).

[13] P. Ginzburg, D. Arbel, and M. Orenstein, IPRA 2005 conference, San Diego, paper JWA6 (2005).

[14] K. Tanaka, and M. Tanaka, Appl. Phys. Let. **82**, 1158-1160 (2003).

[15] E. N. Economou, Phys. Rev. 182, 539 (1969).

[16] I. P. Kaminow et. al., Appl. Opt. 13, 396 (1974).

[17] B. Prade et. al., Phys. Rev. B **44**, 13556-13572 (1991).

[18] R. Zia et. al., J. Opt. Soc. Am. A 21, 2442 (2004).

[19] J. A. Dionne et. al., Phys. Rev. B 73, 035407 (2006).

[20] E. D. Palik (ed.), "Handbook of optical constants of solids", Academic Press (1985).

**Figures:**

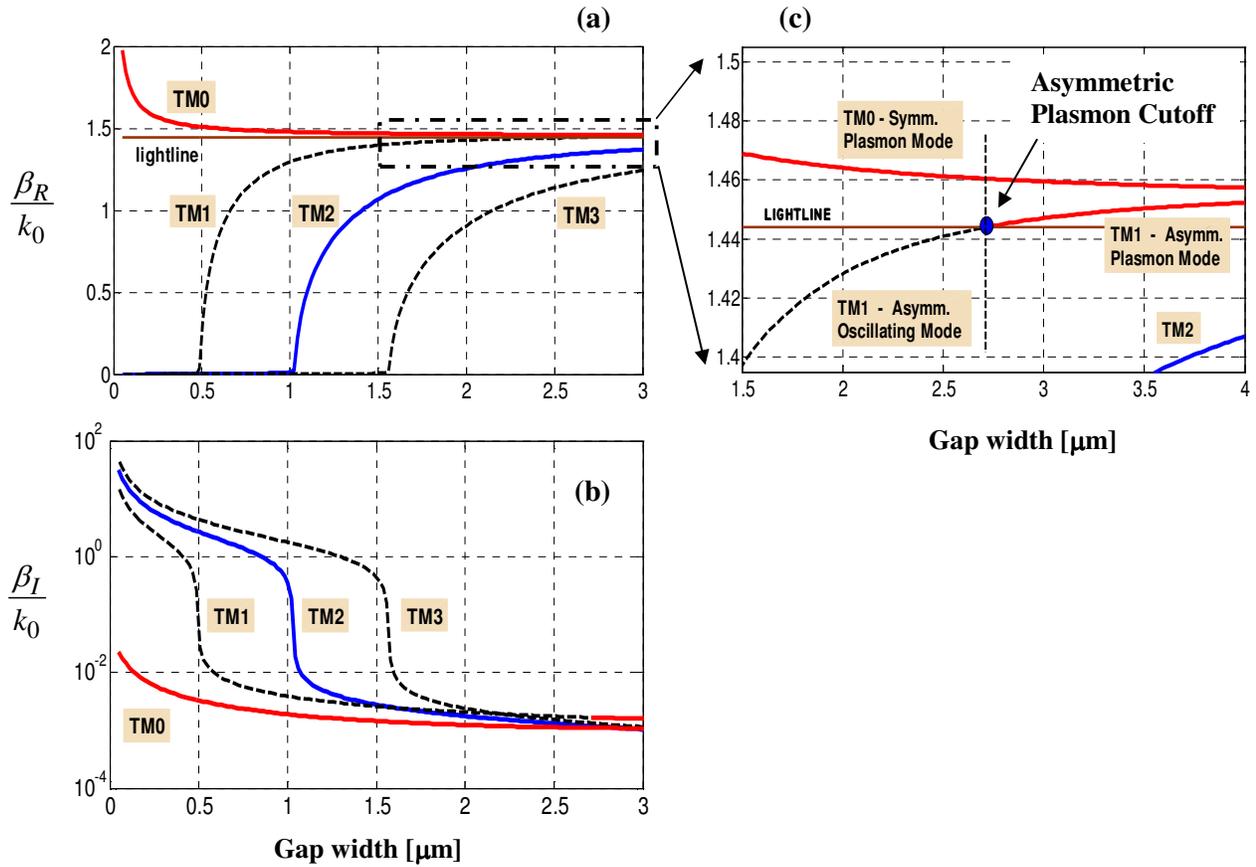

**FIG. 1:** (a) Real part and (b) Imaginary part of propagation constants of the various modes, vs. gap width of PGW. (c) zoom-in on the region of cut-off of the asymmetric plasmon mode.

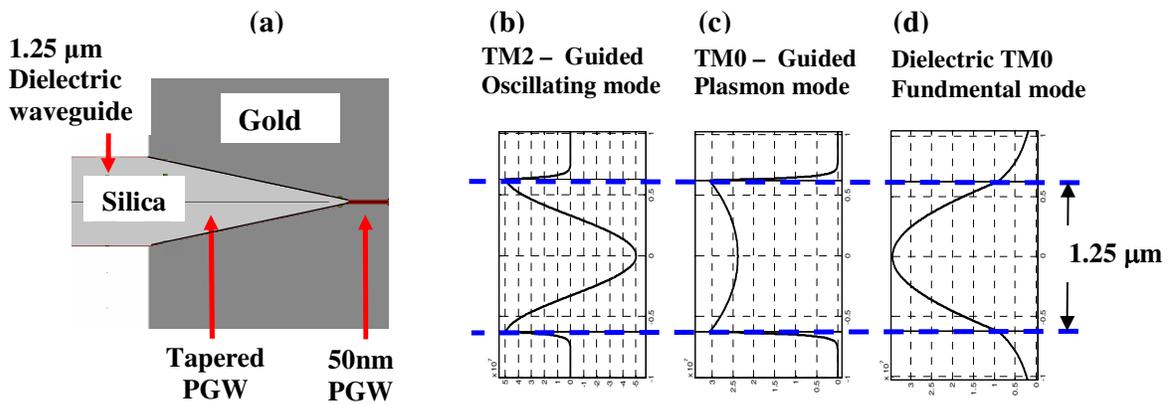

**FIG. 2:** (a) The proposed structure: Tapered Plasmon Gap Waveguide coupler, from a 1.25μm wide input dielectric waveguide, to a 50nm output plasmon gap waveguide.
(b), (c), (d) Magnetic field amplitude ($H_y$) mode profiles: (b) Symmetric oscillating mode (c) Symmetric plasmon mode (d) Symmetric fundamental dielectric mode.

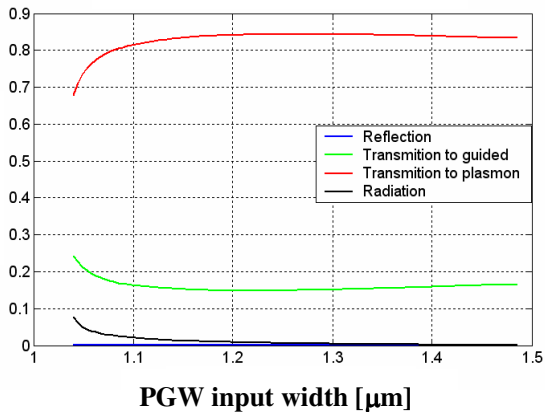 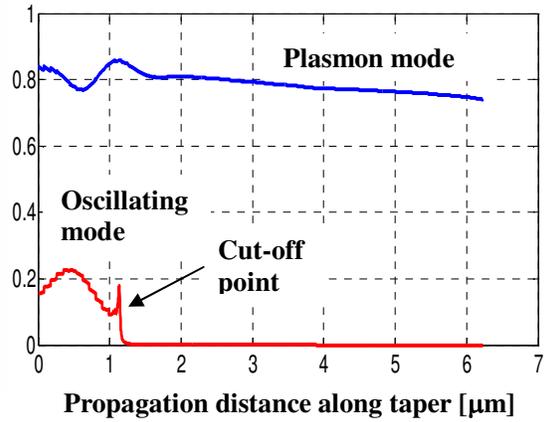

**FIG. 3: (a)** Tapered PGW coupler optimization - coupling of dielectric input mode to the Plasmon and Oscillating modes vs. dielectric waveguide width. **(b)** Intermode coupling and evolution vs. taper propagation distance, for the Plasmon forward/backward and Oscillating forward/backward modes.

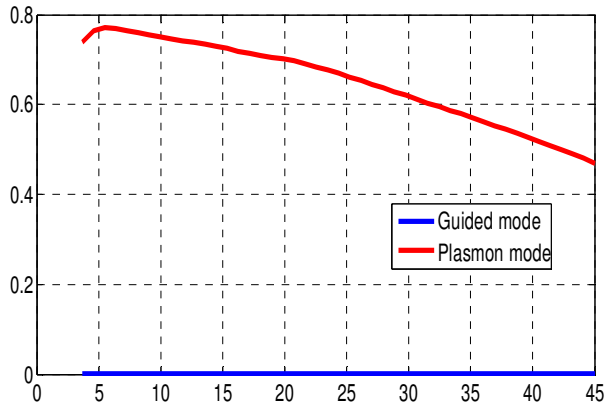

**FIG. 4:** Optimization of taper angle

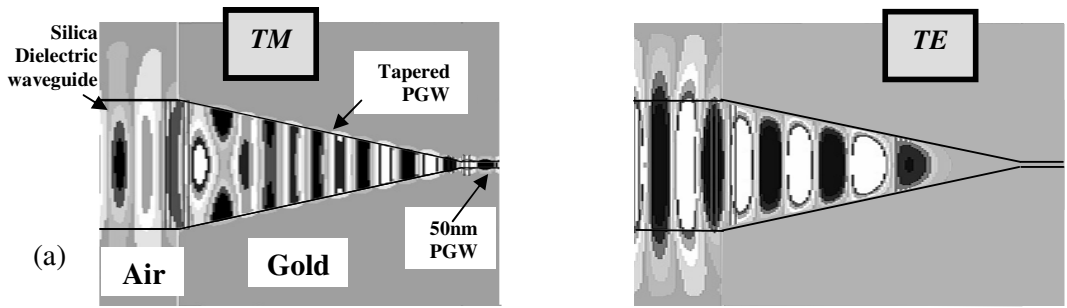

**FIG. 5**: FDTD simulation plot of the tapered PGW coupler. (a) TM field excitation, with >60% transmission. (b) TE field excitation. No plasmon modes exist, and Oscillating mode is cut-off and reflected.